\documentclass[a4paper,fleqn,longmktitle]{cas-sc}

\usepackage[T1]{fontenc}
\usepackage[utf8]{inputenc}

\usepackage[numbers]{natbib}

\usepackage{glossaries}

\makeglossaries

\newacronym{IEA}{IEA}{International Energy Agency} 
\newacronym{NGOs}{NGOs}{non-governmental organizations}
\newacronym{PV}{PV}{photovoltaic}
\newacronym{WEO}{WEO}{World Energy Outlook}

\def\tsc#1{\csdef{#1}{\textsc{\lowercase{#1}}\xspace}}
\tsc{WGM}
\tsc{QE}
\tsc{EP}
\tsc{PMS}
\tsc{BEC}
\tsc{DE}
\begin{document}
\let\WriteBookmarks\relax
\def\floatpagepagefraction{1}
\def\textpagefraction{.001}
\shorttitle{A collective blueprint, not a crystal ball}
%\shortauthors{Göke et al.}
\shortauthors{}
%\begin{frontmatter}

\title [mode = title]{A collective blueprint, not a crystal ball: How expectations and participation shape long-term energy scenarios}                  

\author[1,2]{Leonard Göke}[orcid=0000-0002-3219-7587]
\cormark[1]
\ead{lgo@wip.tu-berlin.de}
\author[3,1]{Jens Weibezahn}[orcid=0000-0003-1202-1709]
\author[1,2]{Christian von Hirschhausen}[orcid=0000-0002-0814-8654]

\address[1]{Workgroup for Infrastructure Policy (WIP), Technische Universität Berlin, Straße des 17.\,Juni 135, 10623 Berlin, Germany}
 \address[2]{Energy, Transportation, Environment Department, German Institute for Economic Research (DIW Berlin), Mohrenstraße 58, 10117 Berlin, Germany}
 \address[3]{Copenhagen School of Energy Infrastructure (CSEI), Department of Economics, Copenhagen Business School, Porcelænshaven 16A, 2000 Frederiksberg, Denmark}

\cortext[cor1]{Corresponding author: \href{mailto:lgo@wip.tu-berlin.de}{lgo@wip.tu-berlin.de}}

\begin{abstract}
The development of energy systems is not a technocratic process but equally shaped by societal and cultural forces. Key instruments in this process are model-based scenarios describing a future energy system. Applying the concept of fictional expectations from social economics, we show how energy scenarios are tools to channel political, economic, and academic efforts into a common direction. To impact decision-making, scenarios do not have to be accurate --- but credible and evoke coherent expectations in diverse stakeholders. To gain credibility, authors of scenarios engage with stakeholders and appeal to the authority of institutions or quantitative methods.

From these insights on energy scenarios, we draw consequences for developing and applying planning models, the quantitative tool energy scenarios build on. Planning models should be open and accessible to facilitate stakeholder participation, avoid needlessly complex methods to minimize expert bias and aim for a large scope to be policy relevant. Rather than trying to simulate social preferences and convictions within engineering models, scenario development should pursue broad and active participation of all stakeholders, including citizens.
\end{abstract}

\begin{highlights}
\item Energy scenarios are tools to channel diverse efforts into a common direction
\item To impact decision-making, scenarios do not have to be accurate --- but credible
\item Scenario creation should combine modeling with broad stakeholder participation
\end{highlights}


\maketitle

% teil zu soziologie und bottom up planning verlängern
\section{Introduction} \label{mot} 

The development of energy systems is not a deterministic process dictated by technology but is equally shaped by societal and cultural forces. How systems develop is closely interrelated with the socially contingent visions of the future that determine how actors direct and coordinate their efforts. A particularly illustrative example to demonstrate this is the evolution of renewable energy over the last 200 years.

First visions for renewable energy systems surfaced in utopian literature authored in the early stages of industrialization \citep{Armytage1956}. In 1865 William Stanley Jevons' \textit{The Coal Question} warned about the expected depletion of coal reserves sparking one of the first public debates on energy supply \citep{Ergen2015}. The idea of renewable energy took a deeper hold and in the late 19\textsuperscript{th} century advances in solar and wind power attracted academic attention rendering them potential substitutes for steam engines \citep{Kapoor2019}. In his main work \textit{Women and Socialism} August Bebel predicted that after the depletion of coal, a shift to renewable energy was inevitable and would lead to a valuation of land based on renewable potential \citep[][cited by \citealp{Abelshauser2014}]{Bebel}. Later Émile Zola's novel \textit{Travail} in 1901 or Archibald Williams' book \textit{The Romance of Modern Invention} in 1910 introduced the idea of renewable energy to a popular audience \citep[][cited by \citealp{Febles2008}; \citealp{Williams1910}, cited by \citealp{Ergen2015}]{Zola1901}.

However, in spite of showing promise, renewable energy systems remained science fiction and the industrial shift from steam to electricity was, with the exception of hydropower, mainly fueled by coal and later oil. Insufficient technical maturity as a sole explanation for this falls short of how technical and societal development are intertwined. Building on Thomas \citet{Hughes1983} historical studies on electricity supply, \citet{Ergen2015} points out that early inventors were too focused on engineering and failed to convey a broader vision for renewable energy systems to governments and private investors. As a result, renewables did not receive long-term investment to achieve learning effects and reach maturity. As an additional reason for little support for renewable energies, \citet{Abelshauser2014} cites that Jevons' expectations regarding the scarcity of fossil fuels turned out to be unfounded, especially due to the increasing exploitation of oil and gas.

Correspondingly, renewables re-gained momentum in the United States when awareness of the risk of depending on oil imports increased during the 1950s. Solar technology progressed at the same time and the presentation of the first photovoltaic cell in 1954 met great public reception being described by the New York Times as the eventual \textit{"realization of one of mankind’s most cherished dreams --- the harnessing of the almost limitless energy of the sun for the uses of civilization"} \citep[][cited by \citealp{Ergen2015}]{NYT}. However, \gls{PV} was not ready for the market and lacked government support that rather focused on advancing nuclear power, also due to synergies between its civil and military use \citep{Clarke2014, Ergen2015}.

In subsequent years, growing opposition to nuclear power and oil dependency reinforced interest in renewable energy systems. An early study on renewable-based energy systems appeared in Denmark in 1975 \citep{Sorensen1975}. In his article \textit{Energy Strategy: The Road Not Taken?} Amory \citet{Lovins1976} contrasted energy systems characterized by fossil and fissile fuels and large-scale infrastructure with an alternative path based on energy efficiency and small-scale renewables tailored to end-use. \citet{Krause1980} substantiated these ideas for West Germany and in 1980 introduced a comprehensive concept for a renewable energy system. Around the same time, similar studies on technical feasibility and economic implications were also conducted for other countries like the United States, France, or Sweden \citep{Martinot2007}. The common denominator in all these works is how they take a systematic perspective to propose serious visions and thus mark the point where renewables exited the realm of science fiction and entered public debate.

Since the 1980s, these visions were partly put into practice. With scientific consensus about climate change as an additional driver, renewable energy made significant progress, quadrupling its global consumption of primary energy from 1980 to 2019. Due to rising consumption in emerging economies, global shares only rose from 6.7\% to 11.4\%. But in industrialized countries, the increase was more pronounced, for example from 1.4\% to 17\% in Germany \citep{owidenergya}. Reinforced investments induced learning effects that diminished levelized costs of electricity from \gls{PV} and wind in the last ten years by 89\% and 70\%, respectively \citep{owidenergyb}.

The trend towards renewables is reflected again by the visions that are projected into the future today. Studies on renewable energy systems are growing in number and detail \citep{Hansen2019,Jacobson2017,Hohmeyer2014,Oei2020,Khalili2022}. While in the 1980s many scholars expressed their skepticism or blunt rejection of the idea, a consensus emerged that renewable energy will make a significant contribution to energy supply and controversy shifted to whether renewables can fully replace nuclear and fossil fuels \citep{Hammond1977,Schmitz1980,Brown2018b}. Outside of academia, new visions for renewable futures, like the Green New Deal, are put forward. Many stimulus packages in response to the COVID-19 pandemic were committed to a green recovery and in reaction to the war in Ukraine many countries reinforce efforts to reduce dependence on fossil fuels, eventually moving renewable energy into the center of public policy \citep{Galvin2020,carbon,wsj}.

This outlined (and non-exhaustive) history of renewables demonstrates how visions of the future drive the energy system. As researchers working on energy scenarios and planning models, the common tool to derive scenarios, we strived for a deeper understanding of these ideational drivers and thus the societal context of our own research. Although there is extensive literature analyzing the creation and use of energy scenarios, none investigates the fundamental societal function of energy scenarios based on socio-economic theory. To close this gap, we investigate how social structure and convictions shape the energy system through scenarios. For this purpose, we interpret energy scenarios as fictional expectations, a sociological concept stating how imaginations of the future drive economic activity. To exhibit how energy scenarios feature the characteristics of fictional expectations, we perform a narrative review that synthesizes theory on economic imaginations in general with research on energy scenarios in particular. Detailing the social contingency of scenarios and how they rather shape than foresee the future, this analysis serves as a basis to derive practical recommendations for working with energy planning models. In conclusion, the paper provides energy experts with a critical perspective on the performativity of energy scenarios and how to consider it within their work.

The following section introduces Jens Beckert’s concept of fictional expectations driving economic development \citep{Beckert2013} and applies it to how the planning of macro-energy systems builds on scenarios. Section~\ref{bot} gives a brief introduction to planning models as a key method for creating scenarios. Section~\ref{zus} links the two previous sections and applies the insights on energy scenarios to derive implications for developing and applying planning models. The last section concludes.

\section{Fictional expectations in energy scenarios} \label{fic}

According to Beckert, economic behavior regarding the future is limited by unknowability. Unlike uncertainty, unknowability is not only non-deterministic, it also cannot be captured probabilistically to enable rational decisions. To remain capable of taking decisions, intentionally rational actors develop fictional expectations of the future in place of perfect information. These expectations extend empirical facts with assumptions that are narratively convincing and formed by calculation and imagination. As a result, fictional expectations are socially contingent and rest on conceptions “influenced by culture, history and power relations” \citep[][cited by \citealp{Jackson2017}]{Beckert2016eng}. Although not inevitably accurate, fictional expectations are treated by actors \textit{"as-if"} and, thus, decisively shape their actions \citep{Beckert2013}. 

Beckert is not the first to recognize the influence expectations and imagination have on human cognition and action. In the fourth century, Augustine of Hippo first reflected on how human imagination forms expectations that only exist in the present, but mentally bridge the gap between present and future, much like memories connecting present and past \citep{Breyfogle1994}. More specifically, other authors have acknowledged an influence on economic decisions. For instance, Pierre Bourdieu described how in Algerian agriculture traditional expectations of the future hindered the transition to a market economy. Accustomed to circular re-reproduction and practical experience, farmers struggled to imagine dynamic growth and abstract institutions, like money, credit, or markets \citep[p.~23]{Beckert2016eng}. Beckert also adds to a more recent line of research that takes a narrative and performative perspective on the economy and economics. For instance, in \textit{Narrative Economics} Robert \citet{Shiller2019} describes how narratives are an overlooked factor driving markets and investment decisions; in \textit{An Engine, Not a Camera} Donald \citet{MacKenzie2016} argues that modern economic theory rather shapes than describes financial markets.

In application of his concept, Beckert revisits microeconomics to explain how fictional expectations open a creative moment for innovation and, more importantly, considers (macro)economic forecasts or projections of technological development as fictional expectations whose societal function is not to predict the future, but to coordinate economic activity \citep{Beckert2016eng}.

In the energy sector, the equivalent of economic forecasts are scenarios.\footnote{In this paper, we use the term \textit{scenario} in a broad sense referring to any imagined future of the energy system. Opposed to that, some sources only use the term for alternatives to a reference development which is labeled \textit{forecast}.} Originating from military planning, scenarios are strategic tools describing a hypothetical future and pathways from the present to this future \citep{Nielsen2007}. Summarizing the process of scenario development, \citet{acatech2016} describe how typically, organizations like governmental agencies, \gls{NGOs}, or companies commission scientific institutions or consulting firms to develop scenarios. In a dialogue the commissioning and the commissioned institution agree on storylines and their translation into numerical assumptions, like fuel prices or demand projections, decide on the modeling approach and define system boundaries for the analysis. The commissioned institution then carries out the analysis using computer models, stylized representations based on economic and physical laws. The interpretation of model results to draw conclusions follows and relates quantitative findings back to the initial storylines. Finally, the scenarios are documented and published, providing two layers of information: First, storylines and the corresponding interpretation of results addressing the general public; second, the underlying quantitative assumptions results and modeling methods addressing other experts.

Scenarios are often categorized as either predictive describing the most probable future, explorative investigating an imaginable future, or normative outlining a desirable future \citep{Nielsen2007}. Correspondingly, Beckert describes economic forecasts as \textit{"coordinating, performative, inventive, and political"} \citep[p.~217]{Beckert2016eng}. Different energy scenarios reflect these properties to a varying degree, depending on the author: Scenarios commissioned by governmental agencies are mostly predictive aiming to coordinate public policy and private investment, but to build credibility and acceptance they refrain from innovation and reflect political consensus. Scenarios launched by \gls{NGOs} or industrial companies seek to influence public policy and opinion in their respective favor. Therefore, they are more political, often normative, and sometimes, when it is inevitable to achieve their objective, innovative. Scientific scenarios exhibit the greatest level of innovation and exploration, especially when strictly addressed to an academic audience and not the general public. In addition, energy scenarios can greatly differ in scope and can range from comprehensive pathways for the entire world, like the IPCC report \citep{IPCC2014}, to explorative concepts for the energy supply of single buildings \citep{Knosala2021}.

To examine how scenarios shape decisions and drive the development of energy systems, the following sections successively transfer the characteristics of economic forecasts identified by Beckert to energy scenarios. For this purpose, each characteristic is briefly introduced and then compared to observations of energy scenarios from the literature. As a result, the section's structure loosely mirrors the chapter on economic forecasts in \citet{Beckert2016eng}.

\subsection{Power of persuasion} \label{per}

To convince recipients and guide their decisions, economic forecasts must be convincing. Beckert identifies two instruments to achieve this, which are typically combined: quantitative model results and narrative elements. For example, economic forecasts consist of a computed growth rate and a story about economic development to support the rate. The purpose of quantitative results and the underlying mathematical method is to evoke precision and objectivity \citep[p.~220]{Beckert2016eng}. To reinforce this perception, method complexity is even increased if it does not benefit accuracy \citep[p.~226]{Beckert2016eng}. Narrative elements convince by suggesting causal relationships and tying economic forecasts to existing knowledge and convictions of the recipients \citep[pp.~91,~221,~245]{Beckert2016eng}.

Energy scenarios combine quantitative methods and narrative elements in a very similar way \citep{Nielsen2007}. Elaboration of quantitative and narrative parts varies and scenarios targeting a professional audience typically emphasize quantitative elements; targeting a general audience emphasizes narratives.

Similar to economic forecasts, the purpose of complex methods in energy scenarios is to create legitimacy and credibility \citep{Schmidt2020}. \citet{Aykut201X} attributes the authority of energy models not only to their academic reputation and seemingly quantitative nature but also to their intransparency to non-experts. This is for example reflected by the critique of the IIASA energy scenarios in \citet{Keepin1984}. This critique, only possible because Keepin worked at IIASA, finds that the applied model is needlessly complex and simple calculations using a few key assumptions are sufficient to largely reproduce its results \citep{Keepin1984,Hafele1981}. In addition, results are not robust to minor changes in these key assumptions. Based on this analysis, \citet{Wynne1984} even goes as far as stating that energy models are \textit{"symbolic vehicles for gaining authority"} only used to create an appearance of scientific objectivity.

The role of narrative elements in energy scenarios is equally acknowledged in the literature \citep{Upham2016, Moezzi2017}. In contrast to purely technical descriptions, stories get people engaged and create awareness of the societal significance of energy scenarios, which is particularly relevant for public policies \citep{Miller2015,Janda2015}. 

Analogously to economic forecasts, energy scenarios are deemed plausible, if narratives correspond to the knowledge and convictions of the recipient \citep{Schmidt2020}. For instance, \citet{Aykut201X} identifies energy scenarios that are based on official economic forecasts and thus avoid controversial assumptions on economic development to be decisive for the success of the German anti-nuclear movement. Conversely, rejection of similar efforts in France is ascribed to putting forward scenarios with rigid assumptions on energy consumption.

To appear plausible and desirable, narratives for transformative scenarios often draw on similar events in the past like the industrial revolution or disruptions in other industries \citep{Janicke2009,Clark2013}. For the same reason, energy scenarios are often associated with various political ideals. Although there are several studies questioning the objectivity and authority evoked by quantitative methods in energy scenarios, there is no research critically examining the association of normative ideals yet. This seems particularly intriguing because associations are often conflicting and range from a broad political spectrum. For example, sovereignty is a reoccurring argument in favor of national resources, both conventional and renewable, and was first cited by Jevon regarding Britain's dependence on coal \citep{Jevon,Sica2016}. Economic growth and the creation of jobs are also equally used as an argument in favor of conventional and renewable energy \citep{whiteHouse}. At the same time, the idea of degrowth serves as an argument for renewables as well. Thermal energy is often promoted arguing it will induce economic growth to lift a significant share of the world's population out of poverty. On the other hand, ambitious mitigation scenarios are often motivated by pointing out how climate change has the most severe impact on people in poverty. Finally, it is argued that renewable energy increases democratic participation through decentralization of energy supply \citep{Krause1980,Hirschhausen2018,Stephens2019}. 

Beyond methods and storylines, the persuasiveness of a scenario also depends on the authority of its author \citep{Schmidt2020}. \citet{Braunreiter2018} for instance observe that researchers refrained from citing a scenario by an environmental organization, not because its quality was questioned, but because it would not \textit{"look serious"}.

\subsection{Coordinating actors} \label{cord}

Beckert points out that despite becoming increasingly sophisticated, economic forecasts are rarely accurate and have a long record of not foreseeing recessions \citep[pp.~223,~241]{Beckert2016eng}. He ascribes this to flaws of the economic forecasting process itself, like incomplete data, inaccurate models, and the inability to foresee major changes but also to unpredictable exogenous shocks to the economy, like political events or natural disasters \citep[pp.~228-231]{Beckert2016eng}. Today, we must add global pandemics and wars to the list, demonstrating how unpredictable exogenous shocks are. Nevertheless, considerable effort is dedicated to forecasting and forecasts receive great interest, because their true purpose, according to Beckert, is to coordinate economic activity. Facing an unknown future, relying on economic forecasts enables actors to be seemingly rational and partly relieves them from responsibility if decisions turn out to be poor \citep[p.~236]{Beckert2016eng}. Consequently, government, businesses, and consumers adapt their decisions to the same forecasts and as a result forecasts achieve consistent behavior across the whole economy. This implies that economic forecasts are performative meaning they do not just predict, but also shape economic activity \citep[p.~237]{Beckert2016eng}.\footnote{It does not necessarily imply that forecasts are self-fulfilling. For example, an optimistic forecast on the eve of an unexpected recession might in fact have the opposite effect and deepen the recession.}

\begin{figure}
	\centering
		\includegraphics[scale=0.8]{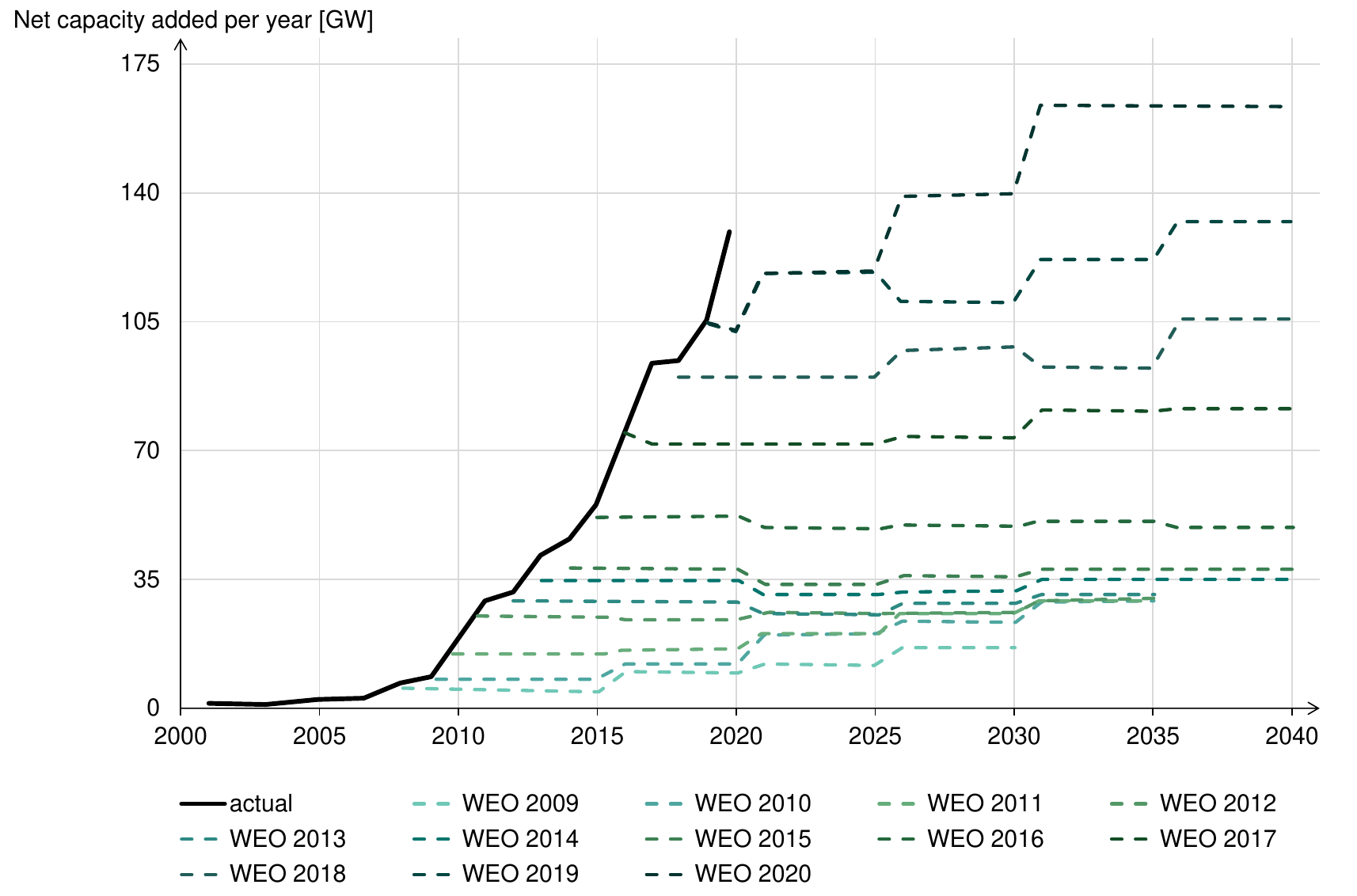}
	\caption[Comparing actual global additions of solar capacities to WEO]{Comparing actual global additions of solar capacities to WEO, data from \citet{weoSolar}}
	\label{fig:weoSolar}
\end{figure}

Again, Beckert's characterization of economic forecasts can be transferred to energy scenarios. Overall, projections of energy scenarios are frequently inaccurate, just as economic forecasts, partly because they must build on economic forecasts for assumptions on energy demand \citep{Paltsev2016,Stern2017,Trutnevyte2016,Nemet2021}. The difficulty to predict how unprecedented events will affect the energy system adds to the inevitable inaccuracies. To stay abreast, even long-term scenarios 30 years into the future, like the \gls{WEO}, are outdated so quickly that they require substantial revision every year. Similar to economic forecasts, energy scenarios struggle to incorporate exogenous shocks affecting demand or fuel prices \citep{Govorukha2020,Craig2002}. In addition, scenarios often neglect major changes and assume an overly conservative continuation of present trends.\footnote{An exception are extreme scenarios used to stress test critical infrastructure, which differ from the long-term scenarios of macro-energy systems considered in this paper.} A prominent example of this bias is displayed in Figure~\ref{fig:weoSolar}. It shows how the \gls{WEO} continuously underestimated solar installations in the last ten years although estimates were continuously revised upwards \citep{Metayer2015}. Although in hindsight continuously missing such developments is baffling, for scenario developers in the moment it can be difficult to assess if such divergence is a temporary deviation or a fundamental long-term trend considering the substantial uncertainty.

Irrespective of accuracy, scenarios impact the development of the energy system, because they affect the decisions of governments and investors \citep{Schubert2015}. This applies in particular to predictive scenarios by established institutions like the \gls{IEA}, large companies, or governments. For example, higher solar projections in the \gls{WEO} could have encouraged additional investments and further driven up renewable expansion \citep{Carrington2018}. Similarly, \citet{Midttun1986} argue that the expansion of French nuclear capacities in the 1970s anticipated increasing electricity demand based on prominent scenarios. Although their underlying assumption on the relationship between economic growth and electricity demand proved wrong, these scenarios became true in a self-fulfilling way, since the state-owned power company instead promoted electric heating to induce the anticipated increase in demand \citep{Aykut201X}.

The importance of credible scenarios for decision making conversely implies that diverging scenarios hinder planning and can stall the development of the energy system \citep{Grundwald2011}. For example, the official governmental objective in Germany was to achieve a renewable share of 65\% in power consumption by 2030, but implemented policies would only have achieved around 55\% \citep{Agora2020b,Oei2019}. In addition, very heterogenous actors advocated for much higher shares up to 100\% and projections for consumption that determine how shares translate into absolute values diverged as well \citep{Gierkink2020, Kendziorski2021}. Overall, the emerging uncertainty not only discourages investment in renewables, but in complementary technologies like storage, electric mobility, or electric heating as well.

Explorative scenarios, often from academia or \gls{NGOs}, rarely have a direct impact on investments, but constitute an avant-garde and drive long-term innovation. If they become sufficiently convincing, their ideas are eventually established and included in predictive scenarios. For instance, first scenarios for renewable energy systems referenced in Section~\ref{mot} were rejected by the established experts at the time and did not encourage great investments \citep{Schmitz1980,Hafele1986}. Also, they appear outdated from today's perspective including no \gls{PV}, no electrification of heating, and exclusive use of synthetic fuels in the transport sector. Nevertheless, these scenarios sparked public discussion and further academic research on renewables finally resulting in significant progress and recognition of renewables by established scenarios. A similar case can be made for the use of hydrogen as an energy carrier, which was already described by Jules Vernes in his 1876 novel \textit{The Mysterious Island} and debated among experts since the 1970s, but only included in governmental scenarios recently \citep{Vernes1875,Bockris2013,govUk,whiteHouse}. 

Finally, scenarios do not only influence governments and investors, but other scenarios as well. To increase the legitimacy of their own work, researchers frequently use assumptions or results from established scenarios as inputs to their own scenarios \citep{Braunreiter2018}. As a result, potential bias transfers to academic scenarios, which is especially critical in case of sensitive assumptions like final demand.

In notable difference from economic forecasts, energy scenarios have become more reflective of their epistemic value \citep{Silvast2020}. According to Beckert, economic forecasts purport to predict the future, despite their long record of inaccuracy. The \gls{IEA}, publisher of the \gls{WEO}, on the other hand, acknowledges that \textit{"there is no single story about the future of global energy and no long-term \gls{IEA} forecast for the energy sector"} \citep{weo} and similarly many academic publications stress the fictionality of energy scenarios. However, this clarification is of no difference, because actors nonetheless treat them as predictions when making decisions. So, when the \gls{IEA} states \textit{"the course of the energy system might be affected by changing some of the key variables"} \citep{weo} to underline the limitations of their work, they are modestly omitting that one variable are energy scenarios like the \gls{WEO} itself.

\subsection{Epistemic participation} \label{eps}

Beckert remarks how successful forecasting is rather a discursive than a technical practice. Economic forecasts for example are not prepared in a purely technical process by a small circle of isolated experts exclusively relying on mathematical models. Instead, forecasters extensively discuss their work with relevant practitioners, like banks, policy-makers and industry representatives, to "calibrate" their models until results appear plausible and consist with the expectations of the stakeholders involved. Thanks to such \textit{"epistemic participation"}, economic forecasts are credible and foster exchange between diverse economic actors coordinating their decisions and stabilizing the economy \citep[pp.~231-235]{Beckert2016eng}. 

Similar to economic forecasts, compelling energy scenarios are not a product of isolated expertise but a discursive process synthesizing the expectations of diverse stakeholders. In this discursive process, scenario developers iterate storylines, input assumptions, and preliminary results with selected stakeholders to achieve consistency with their expectations. For instance, the \gls{WEO} lists more than 100 \textit{"government officials and international experts"} that \textit{"provided input and reviewed preliminary drafts"} \citep{weo}. Often the process draws on workshops to engage experts from various fields and representatives of various interests, for instance in academic scenario development, like the SET-Nav, openENTRANCE, or Sentinel projects, or scenarios by the European transmission grid operators \citep{entso2022,openE}. The latter also launch public consultations to extend feedback on their scenarios.

How energy scenarios project future infrastructure affects citizens' lives in a way much more tangible than the macroeconomic decisions guided by economic forecasts. Lignite, nuclear, or wind power plants shape the immediate environment, electric vehicles alter individual mobility, and above all greenhouse gas emissions warm the earth's climate. For this reason, energy scenarios require broader \textit{"epistemic participation"} than economic forecasts beyond experts and professionals to develop expectations all actors share and will adapt to. 

Accordingly, \citet{Grunwald2019} concludes that scenarios must extend their scope beyond the technical to the socio-technical energy system. Scenarios exclusively reflecting expectations of technical experts and industry pass over an integral part of the system---its user, the citizen. This disregard fueled the debate about nuclear power in West Germany, especially in the 1980s. To experts, on the one hand, the technology was an apparent optimum, an exclusively technical perspective that delegitimized the perspective of many citizens on the other hand, precluding any constructive dialogue.\footnote{Considering social heterogeneity and distributional effects, from a socio-technical perspective a solution can hardly be optimal, but pareto-optimal at best.} To prevent such errors in the future, instead of trying to raise acceptance for predetermined solutions, Grundwald recommends engaging civil society in the early stages of system planning. Recent scenario developments, like the PAC project\footnote{Short for "Paris Agreement Compatible Scenarios for Energy Infrastructure”, see \url{https://www.pac-scenarios.eu/}.} initiated by organizations from civil society, are putting this participative approach into practice.

\subsection{Competition for influence}

Since convincing economic forecasts affect decision-making, their authors hold considerable influence over the future. Therefore, the competition for credibility between forecasters is also a competition for political influence and forecasting methods are assets in this competition \citep[p.~80]{Beckert2016eng}.

In energy, scenarios are key contributions to the debate about the system's future since they shape economic and political decision-making. To influence expectations according to their interests, different actors publish competing scenarios and create a \textit{"battlefield"} of energy system planning \citep[][cited by \citealp{Nielsen2007}]{Norgaard2000}.

For the energy sector, the state holds a central role in this debate, because even in a market economy a large share of planning concerns infrastructure and the environment---both public goods \citep{Midttun1986}. In liberalized power markets, private investors still rely on government owned or regulated grid companies to gain market access. The framework for competition in these markets, partly pre-empting the outcome, is determined by government legislation. For new technologies to emerge, government support, like research funding and subsidies, is a necessary (but not sufficient) condition, as examples like nuclear power or renewable energy demonstrate \citep{Ergen2015}. The availability of fossil fuels and other resources depends on foreign policy and climate legislation. Overall, public policy and long-term government strategy define the framework for private enterprise. Due to their influence on public policy, scenarios commissioned by the government possess considerable authority.

Other actors present in the debate and publishing energy scenarios include industry organizations, environmental \gls{NGOs}, and academia \citep{Kainiemi2020}. It is commonly acknowledged that scenarios often reflect their publishers' interests: for example, scenarios by industry organizations tend towards higher consumption and scenarios by environmental \gls{NGOs} towards lower consumption \citep{Nielsen2007}. Drawing up scenarios must carefully balance between diverging from the consensus to promote own interests on the one hand but complying with the consensus to remain credible on the other. In addition to these professional organizations, the general public, too, increasingly engages in the debate, but unlike other groups, members of the general public do not typically study energy scenarios directly and instead learn about their content from media \citep{Braunreiter2020}.

The role of professional organizations and the authority of complex models restrict influence on the debate and consequently on the development of the energy system. In extreme cases, professional networks between universities, industry, and government form a \textit{"cognitive monopoly"} characterized by a common perspective on the energy system \citep{Midttun1986}. Outsiders pointing out deficiencies of that perspective can be denounced as uninformed by pointing out their lack of recognized expertise or sophisticated methodology \citep{Wynne1984}. Illustrating this, \citet{Midttun1986} describe how environmentalists in several European countries had to establish expertise and forecasting methods of their own to influence public policy according to their interests. At the time, established models and scenarios expected a strong growth in demand that suggested the expansion of nuclear power. In opposition to nuclear power, environmentalists rejected these scenarios and developed methods forecasting constant demand. After gaining recognition, these scenarios were included in the planning of future policies and eventually proved much more accurate. Overall, the process added new perspectives to the debate about energy and created awareness of bias in models.

Concluding from the previous description of how political debate shapes the energy system, a popular but illusive concept of scientific policy advice can be rejected: Decision-makers do not simply derive their actions from the scenarios presented by research, but decisions emerge out of the current state of debate instead \citep{Grundwald2011}. Only scenarios contributing to this debate can have an actual, although indirect impact.

\section{Planning models} \label{bot}

\textit{Bottom-up planning} is the most frequently deployed quantitative method in the development of energy scenarios \citep{Laes2014}. In the past, scenarios from models like MARKAL/TIMES or PRIMES had a considerable impact on energy policy \citep{Taylor2014}. \textit{Bottom-up} modeling refers to representing commodity flows and physical laws in the energy system and is therefore also termed as the \textit{"engineering approach"} \citep{Ringkjob2018}. It is opposed to a top-down approach pursued by integrated assessment or computable general equilibrium models representing cash flows and economic laws. \textit{Planning} implies investigating the optimal design of the system. Planning models are opposed to simulations performed by agent-based or mixed-complementarity models, that account for the individual objectives of different actors. Bottom-up planning models decide on the expansion and operation of technologies to satisfy demand and maximize social welfare. In the case of perfectly inelastic demand, a common assumption, maximizing social welfare is equivalent to minimizing system costs. To solve large and detailed models within a reasonable time, planning models are usually formulated as linear optimization problems. Historically, engineering-based planning models came up as an alternative to econometric and macro-economic models for investigating energy efficiency and alternative energy sources \citep{Aykut201X}.

Planning models are also referred to as partial equilibrium or market models since markets achieve a welfare maximum if perfect competition is assumed. However, these terms can be misleading, because several conditions of perfect competition do not apply to the energy system. First, the unknowability of the future, illustrated by the inaccuracy of scenarios documented in the previous section, violates the assumption of perfect information. Second, the energy system is subject to significant externalities like environmental damage or government intervention. Finally, perfect competition requires a long-term equilibrium, but the transformation of the energy system is a highly dynamic process. Note that violations of perfect competition apply particularly to the expansion of technologies, viz the planning part of models. For welfare maximizing models limited to the operation of predefined capacities the term \textit{market model} can be considered more appropriate.

Thanks to their engineering focus, planning models can identify technically feasible solutions to satisfy demand under certain constraints, for example, an upper limit on carbon emissions. By attributing costs to each decision on expansion and operation, models can also estimate the economic costs of solutions. These characteristics render planning models suitable for techno-economic analyses of energy scenarios and policies \citep{Susser2021}. Typical research includes: investigating the trade-offs between different technologies, for example, heating with electricity or hydrogen; quantifying the benefits of potential innovations, like further decreasing costs of renewables; and fostering a systematic understanding of how technologies interact, for example, \gls{PV} and batteries are typically complements, but power grids and batteries substitute each other \citep{Kendziorski2022}. To analyze specific policies, like a political limit on wind power in certain areas, the effect of the policy must be translated into an appropriate boundary condition of the model, for instance, a corresponding capacity limit.

Since planning models take a system perspective, they do not consider the different agents in the system, like generators, consumers, or regulators and monetary flows between these agents, like subsidies, taxes, or market prices. Accordingly, computed costs are system costs, not the costs of individual agents, and computed prices are opportunity costs of meeting demand, not the market price of transactions between agents. Consequently, planning models are not suited to address strictly economic questions regarding individual profits, market design, or subsidy schemes, which are better addressed by simulative tools. In fact, this is not so much a shortcoming of optimization models but rather reflects how identifying a technical optimum precedes research on implementing the system practically. For example, if planning models robustly find great benefits from expanding electricity storage, but investments are not profitable under current regulations, this is not a flaw of planning, but of the current policy framework.

\begin{figure}
	\centering
		\includegraphics[scale=0.7]{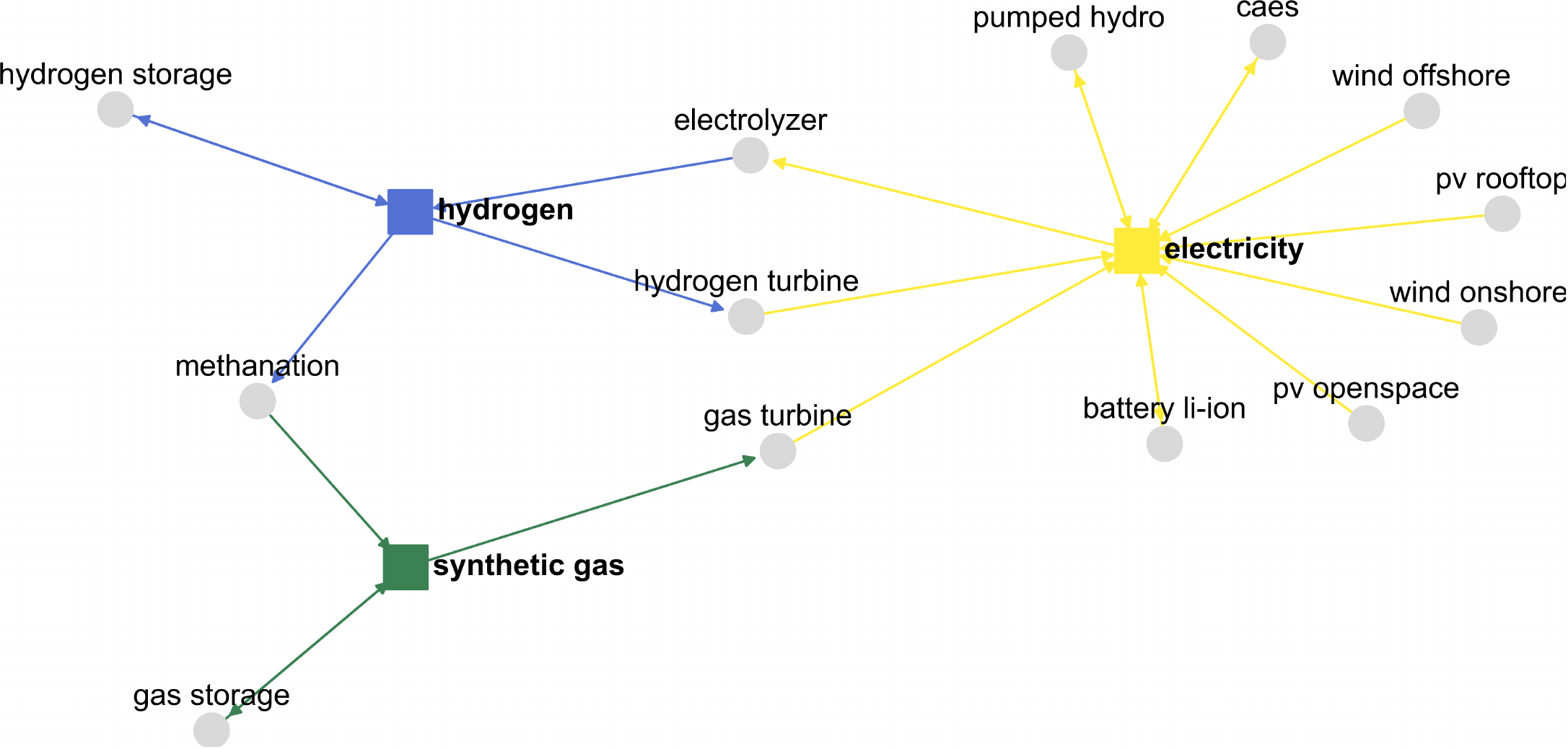}
	\caption{Graph representing exemplary planning model based on \citet{Goeke2021a}}
	\label{fig:ubersicht}
\end{figure}

The scope of planning models ranges from single buildings to countries or continents to the entire world. Since comprehensive decarbonization scenarios have to consider more than one building, the term \textit{macro-energy systems} has been coined to refer to larger macro-systems \citep{Levi2019}. In addition, also the sectoral scope of models varies ranging from only one sector, like the power sector, to coverage of several sectors, like the power, heating, and transport sector, plus their interactions. The graph in Figure~\ref{fig:ubersicht} shows the structure of a stylized planning model. In the graph, carriers are symbolized by colored and technologies by gray vertices. Entering edges of technologies refer to input carriers; outgoing edges refer to outputs. Accordingly, this model is focused on the power sector and additionally includes hydrogen and synthetic gas to represent technologies for long-term storage of electricity.

\section{Implications for planning macro-energy systems} \label{zus}

The creation of energy scenarios heavily relies on planning models and as a result, the societal function of energy scenarios described in Section~\ref{fic} has implications for planning models as well. This section discusses these implications and makes recommendations for further development.

\subsection{Openness and accessibility} \label{open}

Since scenarios are important communication tools in the debate about energy futures, equal opportunity to participate in the debate implies equal access to the scenarios underlying modeling knowledge. The consequences here are twofold:

\begin{enumerate}
  \item First, methods and inputs of scenarios must be transparent, so everybody in the debate has the knowledge to critically assess them. This applies beyond numerical inputs and results. To judge how results reflect the interests of the commissioning organization, normative input assumptions and the concluding interpretation of numerical results must be transparent too \citep{acatech2016}.
  \item Second, modeling tools must be openly available and accessible, so everybody in the debate can contribute scenarios. In the academic literature, the need for both transparent scenarios and open models has been widely acknowledged \citep{Pfenninger2017b,Weibezahn2019,Morrison2018,Junne2019}. Since Keepin's critique of the IIASA model, which was only possible because he worked at IIASA himself, influential models like TIMES have been made publicly available and open-source has become the standard for new models \citep{Wynne1984,TIMES}.
\end{enumerate}

An essential but often overlooked factor in this context, especially beyond the scientific community, is accessibility. Many actors outside of academia do not have the resources in terms of working time or technical knowledge to familiarize themselves with complex data documentations or programming tools. Open modeling tools requiring substantial programming skills and computing resources, scenario data provided in a rare data format, or extensive documentations that are hard to understand all formally comply with openness but do little to open the debate. Similar to how scenario studies address a general public and experts on different layers, open modeling must anticipate different levels of expertise among recipients. The \textit{model.energy} browser application for example provides a stylized but easily accessible planning model.\footnote{https://model.energy/}

Although openness gains popularity, it still faces obstacles. Awareness is high among modelers, but for institutions commissioning scenarios, the designated financiers of openness, it is rarely a priority and without additional budget extensive documentation of code or publication of large data sets is not possible \citep{acatech2016}.

\subsection{Bias minimization}

Analysis in Section~\ref{fic} revealed that quantitative models are not objective tools and are inevitably biased, either by the pursued method, or the assumed parameters. Nevertheless, there are strategies to be transparent about potential bias and minimize it.

Methodologically, the engineering approach of planning models leaves less room for bias than models based on economics or other social sciences. In contrast to economic laws, which can be highly ambiguous---Beckert describes for example how different macroeconomic models either assume a positive, a negative, or no effect of public spending on economic growth---physical laws are unambiguous \citep[p.~229]{Beckert2016eng}. However, even engineering-based models are not exact representations of the physical energy system but must approximate certain laws and heavily aggregate the system to keep complexity reasonable. For instance, models usually approximate the physical power grid by neglecting distribution grids, aggregating the transmission grid into larger nodes, and applying some linear approximation of power flow equations \citep{Leuthold2012}. On the upside, research frequently questions such simplifications and tests them against more accurate representations for validation \citep{Neumann2020,Frysztacki2021,Amirhossein2021}. In addition, an increasing number of studies compares different models to investigate differences and identify bias \citep{Lund2007,Landis2019}. To increase transparency about methodological bias, this research should be pursued further and in addition, scenario studies should openly discuss how their methods might bias results, especially if they diverge from standards.

The second source of bias in planning models are quantitative assumptions. Similar to methods, the reasonable range for technical parameters is much narrower than for parameters that are related to economic or social questions. For instance, calorific values of energy carriers are exactly defined, full-load hours of renewables are limited by empirical data, and Carnot's rule gives maximum efficiencies.

Describing the future, planning models nevertheless require assumptions on technology development that are highly uncertain. Similar to economic forecasts or energy scenarios, technology forecasts are fictional expectations and equally inaccurate \citep[p.~224]{Beckert2016eng}. Technology experts mostly did not foresee the rapid decrease of investment costs for PV or batteries for example. \citet{Bonaccorsi2020} relate these errors to cognitive biases of the experts carrying out technology forecasts, like an anchoring effect of existing forecasts or planning fallacy in development roadmaps. To reduce expert bias in technology assumptions, \citet{Bonaccorsi2020} suggest diversifying participants including various fields and non-experts, nicely reflecting insights on broad participation in energy scenarios from Section~\ref{eps}. \citet{Meng2021} recommend model-based methods to mitigate bias, like Wright’s law, which describes how the expansion of a technology induces learning effects and reduces investment costs. Here, one approach is to include this mechanism, in this case learning rates, directly into planning models \citep{Lopion2019}.

In general, modelers should try to assess parameters critically and refrain from unquestioned adoption of parameters used in other scenarios, even if they hold high authority. All parameters should be documented transparently and key parameters with a strong impact on results, like demand or renewable potential, should be compared to the range from other scenarios and ideally subjected to sensitivity analysis \citep{Wynne1984,Pfenninger2017b}. This is particularly important because, just like any other group of experts, developers of planning models are biased as well. As discussed in Section~\ref{eps}, their \textit{déformation professionnelle} is to have a technical rather than a socio-technical perspective on the energy system favoring technical solutions.

\subsection{Complexity and probabilistic methods} \label{stoch}

Beckert finds little indication that elaborating methods of economic models increased their accuracy and states the actual benefit of refinements is to increase credibility \citep[p.~226]{Beckert2016eng}. For instance, probabilistic methods add complexity and make results harder to refute but are as inaccurate as deterministic methods since the future is unknowable---and not uncertain \citep[p.~43]{Beckert2016eng}. Similarly, \citet{susser2022} investigate if current developments in energy modeling actually improve models and add practical value or just needless complexity.

Against this background, we can draw conclusions on the benefits of probabilistic methods in planning models, an improvement suggested in several publications \citep{Pfenninger2018,Ringkjob2018,Wiese2018}. Considering the limitations of probabilistic methods, they are only recommended for parameters that are actually uncertain, not unknowable. In other words: The probabilistic implementation of a parameter requires a well-founded estimate of its distribution to add accuracy and not just complexity to a model. For example, historic weather data can quantify the distribution of renewable generation very well rendering its probabilistic implementation sensible. On the other hand, deriving a robust stochastic distribution for economic parameters, like capital or investment costs, appears much more difficult, so a probabilistic implementation should be considered carefully and only be pursued to make scenarios more robust---not just for the sake of complexity.

\subsection{Influence on decision making}

Beyond methods and parameters, the ability of scenarios to advance the debate on energy futures decisively depends on how they are deployed. The influence scenarios have on current debates and decisions---often referred to as \textit{policy relevance}---can be used to draw two implications for modeling.

First, scenario scope, and therefore model scope must coincide with the decisions and questions under consideration. Accordingly, \citet{Hughes2013} describe how long-term scenarios for decarbonization rarely have an effect on short-term decisions, because actors find their insights difficult to apply. Overall, scope in planning models can be divided into temporal, regional, and sectoral scope, each having its significance. A temporal scope of multiple years is important to model the dynamics of decarbonization, a large regional scope to consider how energy carriers are exchanged between different regions, and a broad sectoral scope to reflect the utilization of electricity outside of the power sector. For example, an analysis of the additional renewable capacity to achieve a certain renewable share by 2030 has to consider how much of the existing capacity is decommissioned by 2030, the net exchange of electricity with neighboring countries, and the amount of added demand from the heat and transport sector. To cover a large scope while maintaining sufficient technical detail is challenging for planning models. Even if methods reduce computational complexity are deployed, models cannot have all-encompassing scope or detail and must flexibly adapt to questions investigated in a specific scenario \citep{Goeke2020b}.

Second, energy scenarios should carefully consider technical feasibility to prevent severe path dependencies. As introduced in \citet{acatech2017}, the term describes how the number of options to reach a certain objective diminishes with each step towards it. This occurs, if forgone investment in a technology earlier prevents economies of scale and learning effects today or, in the opposite case, if sunk investment costs in the past make a technology favorable today. Adverse path dependencies occur because the future is uncertain, or even unknowable, and past decisions can exclude options preferable in hindsight. In extreme cases, path dependencies can even render the initial objective impossible causing disruption. To demonstrate the disruption, if widely shared expectations, or scenarios, are realized to be impossible too late, Beckert points to the financial crisis of 2008 caused by expectations regarding housing prices and home ownership that suddenly proved false \citep[p.~120]{Beckert2016eng}.

In energy scenarios, speculative assumptions on the availability of technologies, for example, the maturity of technologies like carbon capture and advanced nuclear power or the effective potential of specific renewables, could create similar effects \citep{Braunger2020}. If their availability is widely expected and anticipated by decisions, sudden unavailability will cause failure to achieve mitigation goals. Therefore, scenarios should focus on a risk-averse approach and deviating, more explorative, scenarios should be labeled as such and be aware of imposed path dependencies.

\subsection{Socio-scientific questions} \label{social}

Bottom-up planning takes a technical perspective on energy systems, although Section~\ref{eps} shows how system development is equally a social process. Against this background, a growing number of studies suggest that engineering-based planning models increasingly address socio-scientific questions, too \citep{Niklas2020,Dobbins2020,Krumm2022}. Specific questions to be addressed include: adoption of consumer technologies, acceptance of industry-scale technologies, opportunities for the public to engage, and, similar to past debates described in \citet{Midttun1986}, behavioral change of end-users. Literature provides different proposals on how to address these questions \citep{Dobbins2020,Krumm2022}.

First socio-scientific questions of individual preferences and convictions can be integrated into the mathematical formulation of engineering-based models, but the capabilities of this approach are limited. Behavioral change or attitude towards technology call for simulative or qualitative methods, but since planning models are based on optimization, any representation of social aspects within them is restricted to optimization, too. To address socio-scientific questions anyway, external costs that reflect social aspects can be added to the model and incorporated into the optimization problem, analogously to other cost components. For example, the perceived discomfort from wind turbines can be translated into monetary values and added to the objective function to account for social acceptance of technologies. But the choice of considered externalities and the quantification of their social costs is highly subjective. Estimates for specific technologies often vary by a factor of 100, while estimates for other uncertain parameters, like capital costs, investment costs, or renewable potential rarely vary by more than a factor of two \citep{Stirling1997,FraunhoferEE,Bogdanov2019}. Therefore, the selection of external costs will add substantial bias and easily result in models just reproducing the preferences of modelers.

An alternative to extending planning models is to link them with other simulative tools more appropriate to address non-technical questions. For example, agent-based models can simulate the behavioral change of end-users and based on their results planning models can adjust their exogenous assumptions on demand \citep{Tian2020}. Similarly, the adoption of consumer technologies can be analyzed separately based on models for technology diffusion. Nevertheless, simulative models have well-known shortcomings, for instance, diffusion models are criticized to neglect an adverse impact of innovation on equality \citep{Rogers2003}. As a result, the approach is at the risk of rather adding complexity than value and obscuring bias.

Finally, instead of capturing technology preferences and behavioral factors within planning models or by linking planning models and other tools, citizens, the users of the energy system, can directly participate in the process of developing scenarios and applying planning models. This approach closely relates to the concept of \textit{epistemic participation} outlined in Section~\ref{eps} and reflects that \textit{"models support rather than produce forecasting"} \citep[p.~233]{Beckert2016eng}. Scenarios can include different user preferences, for instance regarding consumer technologies, like electric mobility, and large-scale infrastructure, like wind power plants or transmission lines. Planning models can put a price tag on these preferences by translating them into mathematical constraints. For this purpose, a growing number of studies applies modeling for alternatives with planning models to quantify the change of system costs when deployment of certain technologies is restricted \citep{Buchholz2020,Neumann2020,Pedersen2020,Lombardi2020,Trondle2020b}.

Several studies already investigate active public participation when developing scenarios and applying planning models \citep{Schmid2012,Zivkovic2016}. An example that specifically discusses the role of models is a study by \citet{McGookin2022} that develops energy futures for a rural community in Ireland. The initial approach of the study was to gather concerns and preferences from the community and translate them into an energy planning tool. For instance, due to concerns regarding the affordability of electric vehicles scenarios considered an expansion of public transport. In the course of the study, the scope was broadened even further, because a lot of relevant concerns could not be captured within models at all. Accordingly, the authors conclude from their empirical analysis that a \textit{"narrow focus on adapting energy system models to socio-technical configurations is misguided"}. Instead, they recommend a collaborative and interdisciplinary process to develop scenarios that includes but is not limited to planning models.

In conclusion, striving for holistic planning models to address socio-scientific questions is elusive \citep{Silvast2020}. The strictly techno-economic perspective of planning models is not necessarily a weakness if reflective of its inherent bias and embedded into a participative process between citizens and experts but methods to design such a process are rare \citep{McGookin2021}. They require an interdisciplinary approach that can reconcile the interests of heterogenous social actors with the technical boundaries of the energy system. In return, scenarios will project publicly accepted and thus more robust paths for climate and energy policy.

\section{Discussion and conclusion}

Energy scenarios are a driving force in the development of energy systems and build on planning models for quantitative insights. Against this background, this paper examines how energy scenarios channel political, economic, and academic efforts into a common direction. We find that energy scenarios exhibit all characteristics of fictional expectations: To be persuasive energy scenarios combine narrative elements that get actors engaged and quantitative models that evoke precision and objectivity; the purpose of scenarios is not to forecast the future but to navigate decisions of a diverse range of actors into a common direction. As a result, different scenarios are competing for influence over the development of the energy system. In conclusion, scenarios are communicative tools to shape the energy system and not just purely technical expert projections of the future, like the weather forecast.

Generally, energy scenarios build on engineering-based planning models representing physical flows and investigating the optimal design of the system. Thanks to their engineering perspective, planning models are comparatively robust, but their capability to address certain policy questions, like market design, is limited.

Insights on energy scenarios are applied to draw consequences for developing and applying planning models: The impact scenarios have on the development of the energy system implies that their underlying tools, planning models, should be open and accessible, enabling a diverse range of actors to contribute scenarios and to allow for their critical review. In addition, the impact on system development and public opinion suggests that modelers should minimize and be transparent about the bias imposed by the choice of modeling methods and parameter data. To be relevant for public policy, the spatial, temporal, and sectoral scope of models must coincide with the decisions and questions under consideration. The inability of scenarios to forecast the future suggests focusing stochastic modeling methods to well-quantified uncertainties, like weather-dependent renewable generation, instead of unknowable factors. Rather than trying to simulate social preferences and convictions within engineering models, scenario development should pursue broad and active participation of all stakeholders, including citizens.

In this paper, an economic sociology perspective on energy scenarios was developed and used to derive implications for a specific type of engineering-focused energy models. Future research should adopt a similar approach to reflect on other types of models as well, for instance, simulative economic models capable to address questions of market design. A starting point for such an analysis provides the review of socio-scientific literature on markets and its application to the energy sector by \citet{Silvast2017}. In addition, future research could investigate the performativity of energy scenarios empirically by quantitatively analyzing how modelers create and how recipients use specific scenarios.

\section*{Acknowledgements}

The research leading to these results is part of the OSMOSE project and has received funding from the European Union's Horizon 2020 research and innovation program under grant agreement No. 773406. Also, we want to thank the reviewers for their constructive feedback on earlier drafts of this paper.

\printcredits

%\bibliographystyle{cas-model2-names}
\bibliographystyle{unsrtnat}

\bibliography{cas-refs}

\end{document}